3 May 2024     8 pages, 5 figures

# Astrometric Accuracy of Positions

Erik Høg, Niels Bohr Institute, Copenhagen University

**ABSTRACT:** The great development of astrometric accuracy since the observations by the Greek astronomer Hipparchus about 150 BC has often been displayed in diagrams showing the accuracy versus time. Two new diagrams are provided here, one for positions only, another including parallaxes. New information is included on the Gaia DR3 data and the coming data releases, and about two great astronomers from the distant past, Wilhelm in Kassel and al-Sufi in Isfahan.

1. **New diagrams of accuracy**

The great development of astrometric accuracy since the observations by Hipparchus about 150 BC was documented in 2008 in the first version of the present report. This report was updated in 2017 and again in Høg (2020) with more recent information. The development has often been displayed in diagrams showing the accuracy versus time. Two updated diagrams are provided in the present report as Fig.1 and Fig.1b and by .png files, and this information will presumably be the main interest for some readers. Information on "selected astrometric catalogues" is provided in Høg (2017b) where the values in Table 1 are taken for all ground-based catalogues and for Hipparcos-Tycho.

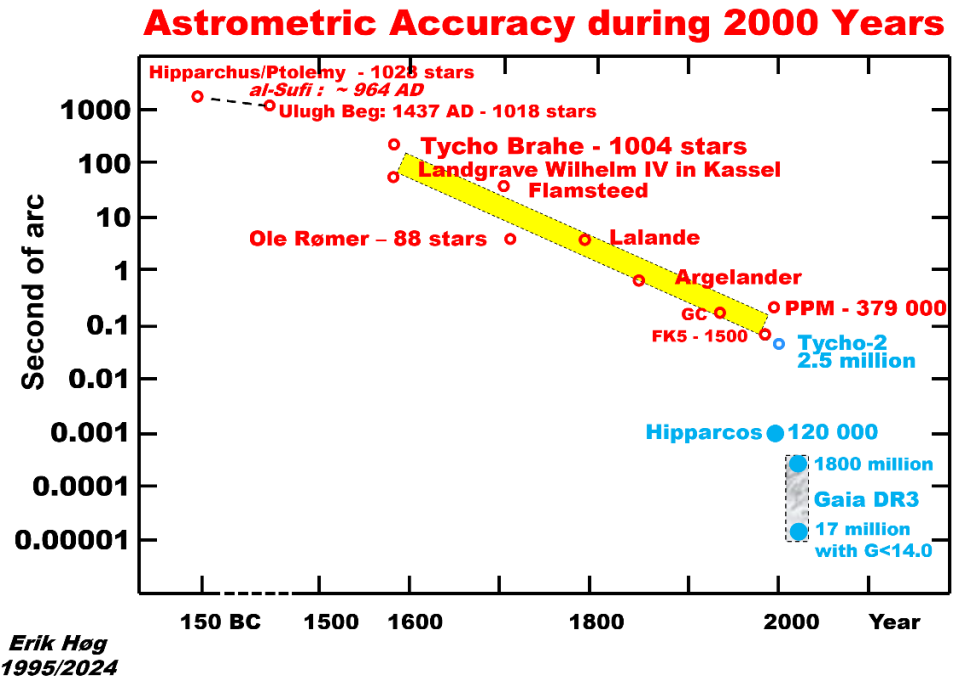

Figure 1. *The accuracy of observed star positions for 2000 years, marked with red color for observations from the ground and with blue from space. The accuracy was greatly improved shortly before 1600 AD by the Landgrave in Kassel and by Tycho Brahe in Denmark. The following 400 years brought even larger but much more gradual improvement before space techniques with the European Hipparcos satellite started a new era of astrometry. Catalogs before 1900 are plotted at the mean epoch of observation, but after 1900 at the time of publication. – The 17 million brightest stars in the sky have accuracies of 0.014 milliarcsec from the first 33 months of Gaia observations. For G < 20.5 mag the accuracy is better than 1 milliarcsecond.*

For simplicity, Fig. 1 is only about positions, omitting parallaxes and proper motions. Catalogs after 1900 are plotted at the epoch of publication, but before 1900 at the epoch of observation. The limit of G<14.0 mag for Gaia was chosen because the uncertainties in Gaia DR3 (Data Release 3) are nearly constant down to this magnitude. Note that I am here using the words *accuracy* and *uncertainty* interchangeably.

The new diagram in Fig.1 from 2024 is available at
https://zenodo.org/records/10977283
and at http://www.astro.ku.dk/~erik/xx/Accu2024.png

Figure 1b is a version of Fig.1 more suited for teaching because it includes ground-based parallaxes. It is like Fig.5 from 1995 in Høg (2008) after a hint from Orlagh Creevey, but it follows our 2024 knowledge about the Landgrave in Kassel, ground-based parallaxes, the Tycho-2 Catalogue and Gaia. Orlagh recently wrote: "In any case, I will indeed use both versions of the figure for all my presentations. … I always like to flash back the huge improvement in astrometry and how this is changing our view of not only the Milky Way but stellar physics too."

The new diagram in Fig.1b is available at
https://zenodo.org/records/10977283
and at http://www.astro.ku.dk/~erik/xx/Accu2024b.png

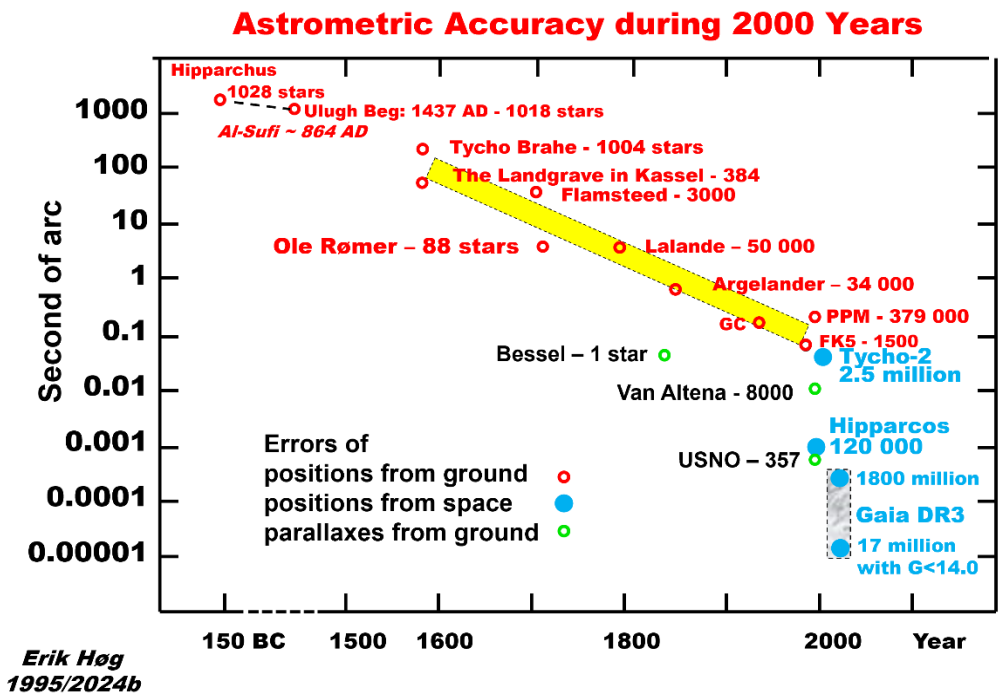

Figure 1b. *Astrometric accuracy, update of Fig.5 from 1995 to our 2024 knowledge, including both parallaxes and positions. Gaia produces millions of accurate proper motions and parallaxes in addition to the positions shown in this diagram, see Gaia DR3 (2022). The important astrometry from space obtained with the HST should also be mentioned, and the very accurate astrometry by interferometry with radio telescopes.*

Figure 1 is derived from the 2019 diagram shown in Fig.2 with the following changes: Gaia DR3 1800 and 17 million stars; only positions, plotted with accuracy better in accordance with Table 1, at the epoch of publication if after 1900, and at mean epoch of observation before 1900; removed some of the number of stars; moved Tycho-2 to the time of publication, and changed the color to blue as for Hipparcos and Gaia; changed to Gaia DR3 median values kindly supplied by Sergei Klioner.

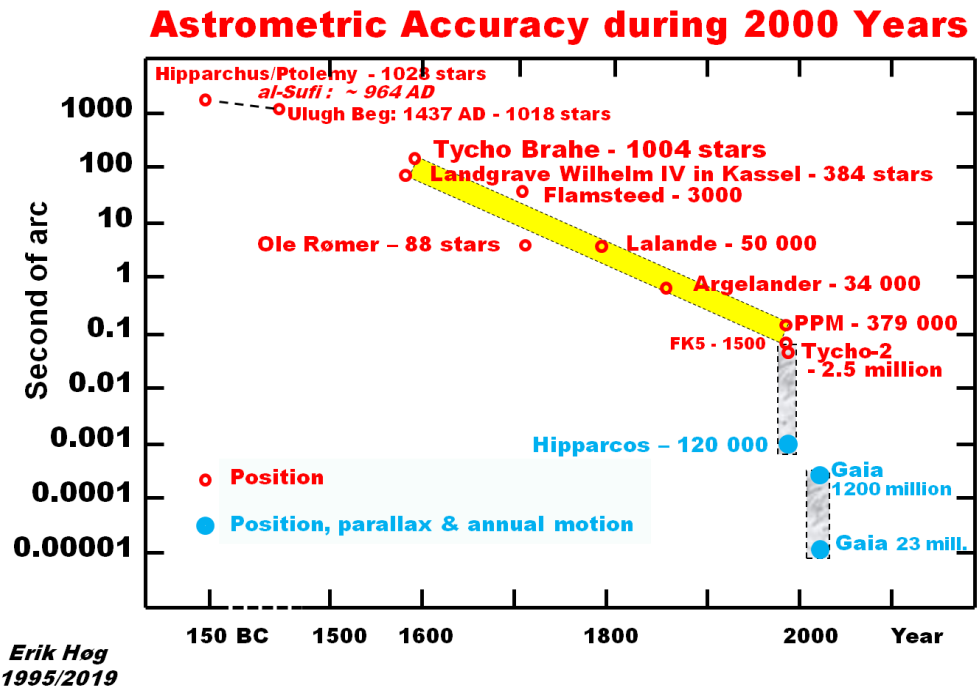

*Figure 2. The 2019 diagram from Høg (2020). It was used as Fig.23 in the review Høg (2024).*

The statistics for Gaia EDR3 with G<14.0 mag, 16844061 sources with median uncertainties: sigma_{alpha*} 0.013 mas where alpha* stands for Δα*cosδ, sigma_delta 0.012 mas and sigma_{pos,max} 0.014 mas, the major axis of the error ellipse.

## 2. Statistics for the diagram in Fig.1

The statistics for DR3 is briefly given above and more follows here from Klioner and Vallenari et al. (2023). Fig.2 of the latter shows the uncertainties of astrometric parameters as function of the G-magnitude. The uncertainty for parallaxes is shown in Fig.3.

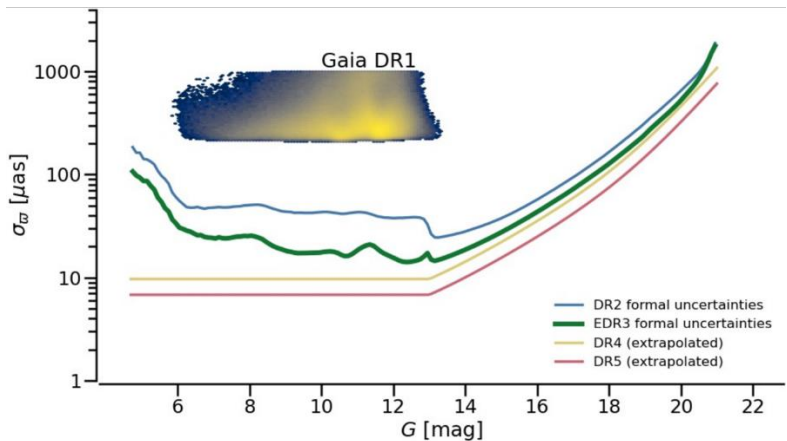

*Figure 3. Uncertainties of parallaxes observed with Gaia.*

It appears to be fairly constant for sources with G<14.0, therefore the uncertainty of this sample of 17 million is plotted in Fig.1. The same figure indicates that the uncertainty for parallaxes is below 1 mas for G<20.5 mag, and we can assume the same for positions.

For the sample of 17 million, Klioner has derived the uncertainties in alpha* and in delta as well as the semi-major axis of the positional error ellipse sigma_pos,max which is computed from the uncertainties in alpha* and delta and their correlations. The medians for DR3 were given above.

The medians for *Hipparcos* were also computed, confirming the values in Table 1 of Høg (2017b)
sigma_{alpha*}  0.87 mas
sigma_{delta}   0.72 mas
sigma_{pos,max} 0.92 mas

But Klioner emphasizes that the distribution is broad! So, median doesn't give all the information. See more on this matter in the following section.

More on the science performance also for the 10-year mission at:
https://www.cosmos.esa.int/web/gaia/science-performance
We can then predict the quality of the resulting *celestial reference frame* far in the future. The proper motion error of DR5 at, e.g., G=20 mag is 0.27*0.325=0.0878 mas/yr. The mean epoch of DR5 will be 2019, the position error 31 years later in 2050 will be 31*0.0878=2.72 mas. In 2080 the error will be 5.36 mas at G=20 mag. - I had this understanding confirmed by Jos de Bruijne at ESTEC.

## 3. Variations of the accuracy

The values in DR3 are from only 33 months of data. The uncertainties will be 1.4 times better in DR4 (66 months of data) and 2 times better in DR5 (126 months of data), an extrapolation with 1/sqrt(interval of observations). The sky coverage will improve with the longer periods of observation. In DR3 some regions in the sky still have rather few observations per stars, see Fig.4, but the uniformity will be much better with time. The present official information by March 2024 about the time of these releases is: Gaia DR4 not before the end of 2025, and Gaia DR5 not before the end of 2030. The Gaia data release scenario is here:  https://www.cosmos.esa.int/web/gaia/release

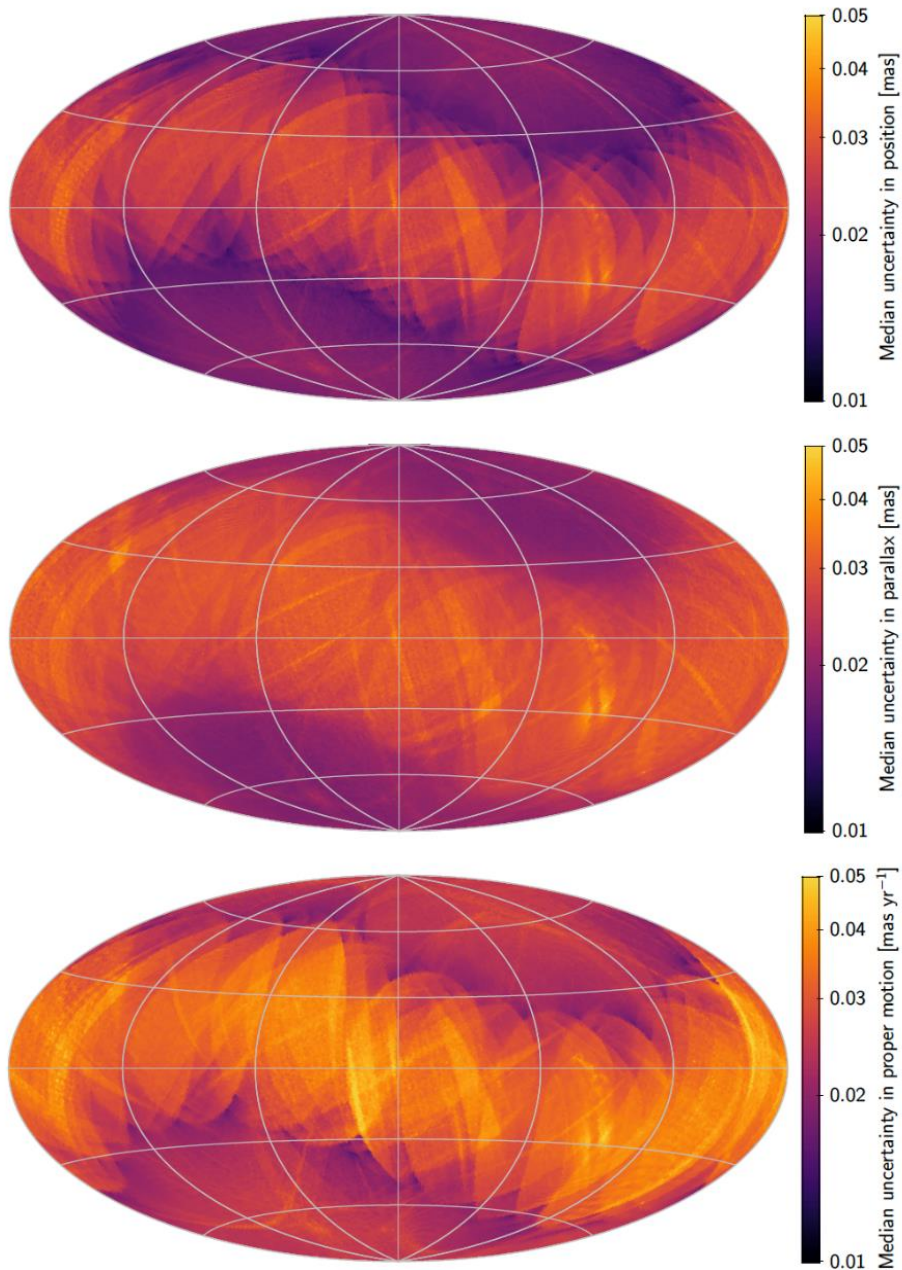

**Fig. 8.** Formal uncertainties at $G \simeq 15$ for sources with a five-parameter solution in EDR3. *Top*: semi-major axis of the error ellipse in position at epoch J2016.0. *Middle*: standard deviation in parallax. *Bottom*: semi-major axis of the error ellipse in proper motion. These and all other full-sky maps in the paper except Fig. 12 use a Hammer–Aitoff projection in equatorial (ICRS) coordinates with $\alpha = \delta = 0$ at the centre, north up, and $\alpha$ increasing from right to left.

*Figure 4. This diagram from a Gaia report shows the variation of accuracy over the sky in DR3. At top, the error of positions appears to vary by more than a factor 4. This variation will be much smaller in later data releases.*

## 4. Other versions of the diagram

Other versions of the diagram are shown in Høg (2008) where Fig.5 shows my original version from 1995.

The version in Fig.8 was shown by Catherine Turon at the IAU Symposium in Shanghai in 2007 as she explained in a mail to me, quoted in the legend. Later, Turon added that *"this graph, originally by Mignard, should be put into its context in Shanghai where it was briefly shown \*only\* for illustration, just to show the drastic improvement provided by space astrometry, not as a careful historical work."*

Now, Catherine wrote: *"Thanks for investigating in depth the numbers of this diagram, usually only used for illustration (as I did in Shanghai in 2007). I have nothing to add."*

An update of this 2007-version is seen as slide #3 in a presentation by Klioner to which Orlagh Creevey sent me a link:
https://indico.strw.leidenuniv.nl/event/2/contributions/5/attachments/13/14/GaiaAstrometricData-Klioner.pdf
Creevey asked for my advice about what to use as she had been informed that I was the originator of the diagram. That sent me into this matter again resulting in the new diagram presented above in Fig.1.

The version in Klioner's slide #3 shows a completely smooth curve for the development of accuracy through 2000 years, and all the catalogs from Hipparchus over Ulugh Bek, Tycho Brahe, Flamsteed, and FK5 to Gaia are placed on his curve. Frankly speaking, I dislike this presentation as unhistorical and misleading. The accuracies of the catalogs do not really lie on a smooth curve, but a *trend* can be seen as the legend to Fig.1 says, and the transition to the two space missions happens not smoothly but in two large *jumps*.

## 5. Notes on the history of catalogs

Due to the very different accuracy given for catalogs and to the very different reliability of such accuracy estimates any other conclusion about the development in time than this trend over 400 years requires great care. Let me conclude with notes on two great astronomers of the distant past, Wilhelm in Kassel and al-Sufi in Isfahan.

**Wilhelm in Kassel**
One of the best experiences for me in these studies was to understand how important the Landgrave in Kassel and the work around him was for Tycho Brahe's life and work. I have explained that in two short papers in English and a detailed one in Danish, Høg (2016, b, c). This insight began at a visit in November 2015 to Andreas Schrimpf and his institute in Marburg to give a talk as I explain extensively on p.8 in Høg (2017a).

Incidentally, Andreas Schrimpf also said that I could write Landgrave in Kassel or Wilhelm in Kassel instead of the more commonly used Landgrave of Hesse. I prefer the first term because his castle lay in Kassel where he lived and made his observations from a balcony, Kassel being a city in the no longer existing county Hesse-Kassel. It is also preferable because the city Kassel is better known to most readers than Hesse which has been the name of many different regions in the course of time.

Accuracies of catalogs are given in Table 1 of Høg (2017b) to the best of my knowledge at the time of writing, but this is a delicate matter. A recent paper by Verbunt & Schrimpf (2021) analyzes the catalogs of Wilhelm IV in Kassel and of Tycho Brahe in detail. I have recently asked the authors if the numbers in Table 1 ought to be revised. Frank Verbunt answered that Table 1 was OK, but that the number 1.14 arcmin for Wilhelm might as well be 1.0 because of the uncertainty in this number.

**Abd al-Rahman al-Sūfī**

Mattia Vaccari has sent me interesting information about al-Sufi who is shown in my figures for his great contributions to astronomy. He lived at the court of Emir Abud al Dawla in Isfahan, and worked on translating and expanding ancient Greek astronomical works, especially the *Almagest* of Ptolemy.

His Wiki is here: https://en.wikipedia.org/wiki/Abd_al-Rahman_al-Sufi

A 2010 PhD thesis claims to be the first one to translate large portions of his original writing: https://researchonline.jcu.edu.au/28854/8/JCU_28854_Hafez_2010_thesis.pdf
I notice that Danish astronomer Hans Schjellerup working at the Copenhagen Observatory translated al-Sufi. The introduction to the thesis says: "The last translation of this significant work was done in French by Hans Karl Frederik Schjellerup back in A.D. 1874 (Schjellerup, 1874)."

## 6. Acknowledgements

I am grateful to Orlagh Creevey for instigating me to review the diagram, and to Uli Bastian, Jos de Bruijne, and Sergei Klioner for discussion and information. It is a pleasure to thank for comments to a previous version of this report Claus Fabricius, Svend Laustsen, Emil Charlie Lind-Thomsen, Andreas Schrimpf, Catherine Turon, Mattia Vaccari, Frank Verbunt, and Volodia Yershov.

Findes i nummer: Kvant 3, 2016

Høg E. 2017a, Astrometric accuracy during the past 2000 years – Partly outdated in 2020
http://arxiv.org/abs/1707.01020v1

Høg E. 2017b, Selected Astrometric Catalogues.
http://www.astro.ku.dk/~erik/xx/AstrometricCats2017.pdf
and at http://arxiv.org/abs/1706.08097

Høg E. 2020, Astrometric accuracy during the past 2000 years.
Placed on arXiv: http://arxiv.org/abs/1707.01020

Høg E. 2024, A review of 70 years with astrometry.
Research Memoir, Invited review for Astrophysics and Space Science. 32 pp. 23 figures
https://rdcu.be/dzp2o

Vallenari et al. 2023, Gaia Data Release 3 Summary of the content and survey properties.
A&A 674, A1 (2023) https://doi.org/10.1051/0004-6361/202243940
or at: https://www.aanda.org/articles/aa/pdf/2023/06/aa43940-22.pdf

Verbunt, F., Schrimpf, A. 2021, The star catalogue of Wilhelm IV, Landgraf von Hessen-Kassel. Accuracy of the catalogue and of the measurements. A&A vol.649, https://doi.org/10.1051/0004-6361/202140332